\documentclass[a4paper]{article}

\usepackage{INTERSPEECH2022}

\usepackage{xcolor}
\usepackage{url}
\usepackage{booktabs,threeparttable}
\usepackage{multirow, makecell}
\usepackage{anyfontsize}

\newcommand\blfootnote[1]{%
  \begingroup
  \renewcommand\thefootnote{}\footnote{#1}%
  \addtocounter{footnote}{-1}%
  \endgroup
}

\title{Simple and Effective Unsupervised Speech Synthesis}

\name{Alexander H. Liu$^{*1}$, Cheng-I Jeff Lai$^{*1}$, \\
Wei-Ning Hsu$^2$, Michael Auli$^2$, Alexei Baevski$^2$, James Glass$^1$\thanks{$*$ Equal contribution}}
\address{
  $^1$MIT CSAIL 
  $^2$Meta AI}
\email{\{alexhliu,clai24,glass\}@mit.edu}

\begin{document}

\maketitle
\begin{abstract}
We introduce the first unsupervised speech synthesis system based on a simple, yet effective recipe.
The framework leverages recent work in unsupervised speech recognition as well as existing neural-based speech synthesis.
Using only unlabeled speech audio and unlabeled text as well as a lexicon, our method enables speech synthesis without the need for a human-labeled corpus.
Experiments demonstrate the unsupervised system can synthesize speech similar to a supervised counterpart in terms of naturalness and intelligibility measured by human evaluation.
\end{abstract}

\blfootnote{{Demo: {\fontsize{6.8}{6.8}{\url{https://people.csail.mit.edu/clai24/unsup-tts/}}}}}

\section{Introduction}

With the recent advance of deep learning, neural-based text-to-speech (TTS) systems have closed the gap between real and synthetic speech in terms of both intelligibility and naturalness~\cite{shen2018natural}.
However, a sizable dataset composed of speech-text pairs is necessary to synthesize high-quality speech~\cite{chung2019semi}.
Consequentially, speech synthesis systems are not available for the vast majority of languages~\cite{tan2021survey}.

In this preliminary study, we make the first attempt (to the best of our knowledge) to achieve unsupervised speech synthesis with the goal of addressing the aforementioned limitation of speech synthesis.
We simulate an extreme situation where human-annotated speech is unavailable by considering only unpaired audio, unpaired text, and a grapheme-to-phoneme lexicon.
We propose a simple two-step recipe to build TTS systems under such conditions: first, utilizing an automatic speech recognition (ASR) model to provide pseudo-labels for untranscribed speech; second, training a TTS model with machine-annotated speech only.

For the first step, we take advantage of recent work in unsupervised speech recognition~\cite{baevski2021unsupervised}.
We train wav2vec-U 2.0~\cite{liu2022towards}, an ASR model that does not require paired speech and text data, to obtain pseudo-speech-to-text annotation.
To build speech synthesize systems as the second step, we follow the same learning paradigm as existing supervised TTS models~\cite{wang2017tacotron}, but use machine-annotated speech instead of human-annotated data.

In our experiments, we demonstrate the effectiveness of the simple method.
We show that synthesizing intelligible and natural speech is possible without the need for human-labeled data by performing both objective and subjective tests.
We also show that unsupervised TTS can perform on par with supervised TTS despite learning from imperfect transcriptions. 
\begin{figure*}[ht]
\centerline{\includegraphics[height=6.15cm]{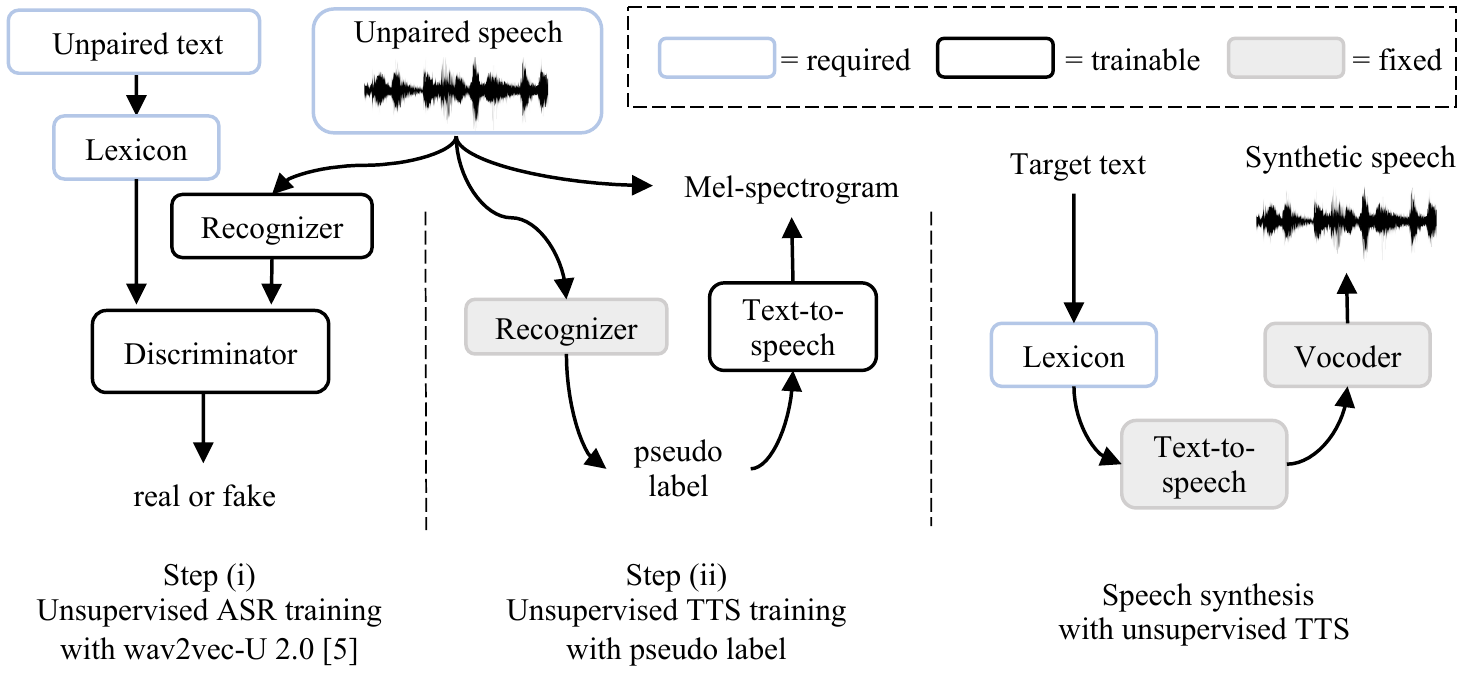}}
\caption{An overview of the proposed framework for unsupervised speech synthesis. Only unpaired text, unpaired audio, and a lexicon are required. Step (i): build unsupervised ASR with wav2vec-U 2.0~\cite{anonymous2022}. Step (ii): train unsupervised TTS with unpaired speech and pseudo labels from the previous step. 
Speech synthesis can be done combining lexicon and vocoder (trained with unpaired speech).
\vspace{-10pt}
}
\label{fig:overview}
\end{figure*}

\section{Background}
\label{sec:related}

\subsection{Supervised Speech Synthesis}

While there are different methods for speech synthesis, we focus on neural network-based TTS systems where the text-to-speech mapping is modeled by deep neural networks using encoder-decoder architectures~\cite{wang2017tacotron,tachibana2018efficiently,li2019neural}.
Under the sequence-to-sequence encoder-decoder paradigm, the input text is first converted into a phone sequence with the aid of a lexicon, and is then encoded into latent features using the encoder.
The auto-regressive decoder predicts Mel spectrograms based on the encoder features.
The entire model can be trained by minimizing the reconstruction error with the target Mel spectrogram.

Besides the text-to-spectrogram models, neural vocoders have also played an important role in neural-based TTS systems~\cite{yamamoto2020parallel}.
Vocoders aim to generate a waveform from synthetic Mel spectrograms by learning from $<$waveform, Mel spectrogram$>$ pairs collected from audio data.
Combining text-to-Mel spectrogram TTS models and neural vocoders, high-quality speech that is indistinguishable from real data can be synthesized ~\cite{van2016wavenet,shen2018natural}.
Nevertheless, considerable amounts of human-annotated speech are required for high-quality neural-based TTS systems. 

\vspace{-6pt}

\subsection{Semi-supervised Speech Synthesis}

To improve the data efficiency for neural-based TTS systems, methods utilizing unpaired training data have thrived.
Unlike paired  $<$audio, text$>$ data, collecting unpaired text or unpaired audio is relatively straightforward.
Prior works found that semi-supervised learning using unpaired data jointly with paired data can reduce the need for paired data in different ways.
For example, pre-training the encoder/decoder~\cite{chung2019semi}, improving input text representation~\cite{wang2015word,jia2021png}, and data augmentation using the opposite natural of ASR and TTS~\cite{tjandra2017listening,ren2019almost,liu2020towards}.
However, existing semi-supervised methods are still bounded by the amount of paired training data~\cite{chung2019semi,ren2019almost,liu2020towards}.
In this work, we push the limits of neural-based TTS systems by using \textit{no} paired training data.

\vspace{-6pt}

\subsection{Unsupervised Speech Recognition}
\label{related:asr}

Since the key to achieving unsupervised speech synthesis with the proposed method is to generate transcriptions of speech without paired  $<$audio, text$>$ data, we first review prior work on unsupervised speech recognition.
Wav2vec-U~\cite{baevski2021unsupervised} achieves good speech recognition performance without paired data by using self-supervised speech representations~\cite{baevski2020wav}.
The method first embeds the unlabeled speech data with wav2vec 2.0~\cite{baevski2020wav} and then determines segment boundaries by clustering the wav2vec 2.0 representations. 
Next, the method builds segment representations by performing PCA dimension reduction on wav2vec 2.0 representations and then pools the features of all time-steps in a particular segment.
These representations are then input to a generator to perform adversarial training.
Text data is also pre-processed via a grapheme-to-phoneme transformation together with silence padding around each sentence and randomly inserted silences between words.
Finally, adversarial training obtains a speech recognizer by mapping the pre-processed features to a phone sequence.
Here, we use an improved version, wav2vec-U 2.0~\cite{liu2022towards} which trains the recognizer directly on the raw audio without requiring special pre-processing of the audio (Fig.~\ref{fig:overview}), i.e., by removing the need to determine segment boundaries and pre-processing the wav2vec 2.0 features.
The main change to wav2vec-U is to output phone predictions with an increased stride which enables removing much of the preprocessing.

\section{Unsupervised TTS}

\subsection{Problem formulation}

As the first study aimed at unsupervised TTS, our goal is to synthesize speech with the following resources:
\begin{itemize}
    \item An audio corpus containing speech without paired text.
    \item A text corpus containing sentences where no exact match exists with the spoken corpus. Furthermore, there is no domain mismatch between the text and audio corpora. 
    \item A lexicon providing the pronunciation representation (i.e., phone sequence) of each word in the text corpus. 
\end{itemize}

We present a pipeline method for building unsupervised TTS with these resources. 
The training procedure is broken into two steps, described in the following subsections.

\subsection{Pseudo labeling speech via unsupervised ASR}

As illustrated in Fig.~\ref{fig:overview}(i), the first step of the proposed method is to generate pseudo-labels for each utterance in the spoken corpus.
To this end, we first train wav2vec-U 2.0~\cite{liu2022towards}, an existing unsupervised speech recognition method described in Section~\ref{related:asr} and detailed in Section~\ref{setup:asr}, on unpaired audio and text.
After training, the audio corpus can be decoded with the resulting recognizer to obtain pseudo-labeled phone sequences.

\subsection{Unsupervised text-to-speech with pseudo label}

For the second step, we train a sequence-to-sequence TTS with the pseudo-labeled audio corpus as shown in Fig.~\ref{fig:overview}(ii) and detailed in Section~\ref{setup:tts}.
The goal of the TTS module is to learn to recover Mel spectrograms that contain spoken content specified by the imperfect input phone annotation.
During testing, Mel spectrograms can be synthesized by feeding the phone sequence representation of the desired sentence.
To generate audible speech, a separate vocoder is used to convert Mel spectrograms into waveforms.  This last step is inherently unsupervised since no text is needed for training a Vocoder.

\section{Experimental Setup}
\label{sec:setup}

\noindent\subsection{Datasets}\label{subsec:dataset}

\noindent\textbf{Audio:}
Two speech synthesis scenarios are considered:
\begin{itemize}
    \item \textbf{Single speaker}: {LJSpeech}~\cite{ljspeech17} contains about 24 hours of read speech from a single female speaker. 
    Following prior work on weakly-supervised TTS~\cite{ren2019almost,liu2020towards}, 300 utterances are randomly selected for both validation and test, leaving 12,500 utterances for training.
    \item \textbf{Multi-speaker}: {LibriTTS}~\cite{zen2019libritts} is a subset derived from LibriSpeech~\cite{panayotov2015librispeech} containing over 500 hours of speech collected from 2,456 speakers. 
    We used the \texttt{\footnotesize{train-clean}} subsets combining roughly 250 hours for training.
    Since unseen speaker modeling is not our focus, we followed the original work~\cite{zen2019libritts} and randomly selected 6 speakers (3 male and 3 female with IDs {\footnotesize\texttt{8699, 4535, 6209, 6701, 3922, 3638}}) from the training split and use text from \texttt{\footnotesize{test-clean}} for testing.
\end{itemize}

\noindent For unsupervised ASR, the audio is downsampled to 16khz to extract speech representations from wav2vec 2.0~\cite{baevski2020wav} to serve as the input.
For TTS, target Mel spectrograms are extracted from the audio (with silence removed) with 80 log Mel filter banks.

\vspace{4pt}

\noindent\textbf{Text:} The official text corpus provided by LibriSpeech~\cite{panayotov2015librispeech} is used as the unpaired text source. Transcriptions of utterances from LJspeech/LibriTTS are excluded\footnote{ \scriptsize\url{https://github.com/flashlight/wav2letter/blob/main/recipes/sota/2019/raw_lm_corpus/README.md}} to ensure there are no matching sentences between the text and audio data.

\vspace{4pt}

\noindent\textbf{Lexicon:}
A word-to-phone mapping is obtained through an off-the-shelf phonemeizer~\cite{g2pE2019}.
To phonemize the text source, all punctuation marks are discarded and the difference between variants of the same phone is ignored.
For our English experiments, this results in a phoneme inventory of size 39 that is used for both ASR and TTS.

\noindent\subsection{Unsupervised ASR model}
\label{setup:asr}
\noindent\textbf{Training:} As illustrated in Fig.~\ref{fig:overview}(i), training wav2vec-U 2.0~\cite{liu2022towards} is the first step for the proposed method.
Under the adversarial learning framework, the goal of the recognizer is to transcribe a speech representation into a phone sequence that is indistinguishable from the real phone sequence to a discriminator.

Taking a 1024 dimensional speech representation as input, the recognizer is a 3-layer neural network consist of a  batch normalization layer, followed by a linear projection, and a convolution neural network.
The recognizer predicts 1 phone for every 9 input frames with a stride of 3.
To change the model output from frame-synchronized to phoneme-synchronized, consecutive repeated phone predictions will be trimmed by randomly preserving one frame.
Besides the adversarial objective, the recognizer is also regularized by output smoothness penalty and phoneme diversity loss~\cite{baevski2021unsupervised}.
The discriminator is composed of a 2-layer convolutional neural network with a receptive field of size 9.
The input is either one-hot vector sequences representing phone sequences from the unpaired text or probability prediction.
Besides the adversarial objective, gradient penalty~\cite{gulrajani2017improved} is also imposed on the discriminator.
We use the fairseq~\cite{ott2019fairseq} implementation for training with default hyper-parameter\footnote{\scriptsize\url{https://github.com/pytorch/fairseq/tree/main/examples/wav2vec/unsupervised}}.

\vspace{2pt}

\noindent\textbf{Decoding:} To transcribe the unpaired audio, each utterance is decoded by the recognizer together with a phoneme-to-phoneme weighted finite-state transducer ~\cite{mohri2002weighted} to remove silence prediction and incorporate a phone-based 6-gram language model trained from the text corpus.
A beam search with beam size 15 is used for decoding with unsupervised hyper-parameters selection~\cite{baevski2021unsupervised}.
As a reference, decoding the test set with wav2vec-U 2.0 results in 6.97\%/7.78\% phone error rate on LJSpeech/LibriTTS respectively.

\noindent\subsection{TTS model}
\label{setup:tts}
\noindent\textbf{Text-to-Speech:}
A Transformer-TTS~\cite{li2019neural}, a sequence-to-sequence encoder-decoder model, is selected as the phone-to-Mel spectrogram model in our framework.
The Transformer-TTS consists of an encoder, an auto-regressive decoder, a pre-net, and a post-net. 
The encoder and decoder have 6 layers of transformer blocks.
We used ESPNet-TTS~\cite{hayashi2020espnet} for model training with an L2 reconstruction error based on its default configurations\footnote{\scriptsize\url{https://github.com/espnet/espnet/tree/master/egs2/}}.
On LJspeech, we found that it is necessary to enforce guided attention loss~\cite{tachibana2018efficiently} in all 48 decoder attention heads. 
Models are trained for 600/1,000 epochs on LJSpeech/LibriTTS respectively.
Besides the Transformer-TTS, we found that Tacotron2~\cite{shen2018natural} has slightly worse synthesis quality and Fastspeech2~\cite{ren2020fastspeech} has issues synthesizing longer utterances. 
We hypothesize that these difficulties arise due to the use of a reduced phone set (see Lexicon in Section~\ref{subsec:dataset}), which makes the text to Mel spectrogram mapping much more difficult in the absence of a pronunciation.
Model selection is based on validation loss, and the best 5 checkpoints are decoded and averaged. 

\vspace{2pt}
\noindent\textbf{Speaker modeling on LibriTTS:}
Following prior work~\cite{jia2018transfer,cooper2020zero,xue2022ecapa}, a pre-trained speaker verification model ECAPA-TDNN~~\cite{desplanques2020ecapa} from SpeechBrain~\cite{ravanelli2021speechbrain} is used as the speaker encoder to generate speaker embeddings.
The Transformer-TTS is conditioned on the speaker embedding and Global Style Tokens (GST)~\cite{wang2018style} for multi-speaker modeling~\cite{cooper2020zero,xue2022ecapa}. 
We found that fine-tuning the TTS on each target speaker for 10 epochs significantly improves the synthesis quality.
During testing, an utterance from the desired speaker is randomly selected from the training split as the reference for the speaker encoder.

\vspace{2pt}

\noindent\textbf{Vocoder:} 
A separate vocoder pre-trained on the same corpus converts the synthesized Mel spectrograms into waveforms. 
On LJspeech, we chose Parallel WaveGAN~\cite{yamamoto2020parallel} that was trained for 3M iterations.
On LibriTTS, HiFi-GAN~\cite{kong2020hifi} is selected as it produces the clearest speech compared to MelGAN~\cite{kumar2019melgan} and Parallel WaveGAN. 
All vocoders are publicly available\footnote{\scriptsize\url{https://github.com/kan-bayashi/ParallelWaveGAN}}. 


\noindent\subsection{Evaluation}
To evaluate the proposed unsupervised TTS method, we compare the speech quality of the following settings:
\begin{itemize}
    \item \textbf{{Unsupervised} ({\tt Unsup})}: synthetic speech from the unsupervised TTS model.
    \item \textbf{{Supervised} ({\tt Sup})}: synthetic speech from supervised topline -- the same TTS model trained with ground truth transcription instead of machine annotation.
    \item \textbf{{Natural} ({\tt Nat})}:  speech  from the dataset.
\end{itemize}

\noindent Three subjective and objective measures are conducted:
\begin{itemize}
    \item \textbf{Mean Opinion Score (MOS)}~: quantifies subjective naturalness, where workers are asked to rate each utterance on a 5-point scale (with 1-point increment).
    100 unique utterances respectively from LJspeech and LibriTTS test sets are selected randomly and designated as the MOS test sets.
    150 HITs (crowdsourced tasks) are run for LJspeech, each HIT includes 10 utterances from each setting.
    200 HITs are run for LibriTTS, each HIT includes 3 utterances from {\tt Nat} and synthesized by {\tt Unsup} and {\tt Sup} for each one of the 6 target speakers.
    
    
    \item \textbf{Word Error Rate (WER)}~: quantifies intelligibility.
    The LJspeech test set (300 synthetic utterances) is fed into Google's ASR API\footnote{\scriptsize\url{https://pypi.org/project/SpeechRecognition/}}
    and we measure the word error rate of the result with respect to the input ground truth.
    
    \item \textbf{Pairwise Opinion Score}~: quantifies subjective pair-wise naturalness and intelligibility (also known as A/B test), where workers are asked to select the more natural or intelligible synthesized utterances between {\tt Unsup} and {\tt Sup}. 
    We ran it for 1000 HITs, and in each HIT, a random utterance is drawn from the MOS test set, synthesized, and presented to the workers.  
    
\end{itemize}

\section{Results}

\subsection{Single Speaker TTS on LJSpeech}

\noindent\textbf{Naturalness:}
Results of evaluating the naturalness are listed in Table~\ref{tab:lj_mos}.
With the upper bound being the 4.05 scored by real data from LJSpeech, we can see the supervised model performed remarkably well by scoring 3.94 with a slightly higher variance.
Compared to the supervised topline, the proposed unsupervised method scored 3.91 with similar variance, degraded by a mere 0.03.
Considering the 6.97\% phone error rate the unsupervised model suffered during training, the small degradation suggests that the proposed pipeline method does not suffer much from error propagation.
Even compared against real speech, the MOS score of unsupervised TTS is only slightly lower.
This demonstrates the unexpected robustness of the proposed two-step pipeline method.

\begin{table}[t!]
\centering
\caption{Speech naturalness measured by  Mean Opinion Score (MOS) with 5-point scale on LJSpeech test split.
\label{tab:lj_mos}}
\begin{tabular}[t]{lcc}
\toprule
\multirow{2}{*}{Method} & Input phone error rate & \multirow{2}{*}{MOS} \\
& suffered during training & \\
\midrule
\midrule
Natural & - & 4.05 $\pm$ 0.07 \\
Supervised & 0\% & 3.94 $\pm$ 0.08  \\
\midrule
Unsupervised & 6.97\% & 3.91 $\pm$ 0.08  \\
\bottomrule
\end{tabular}
\end{table}

\begin{table}[t!]
\begin{threeparttable}[ht!]
\centering
\caption{Speech intelligibility measured by Word Error Rate (WER) with Google ASR on LJSpeech test split. 
\label{tab:lj_wer}}
\begin{tabular}[t]{lcc}
\toprule
\multirow{2}{*}{Method} & Source of input phone & Word error rate \\
&  sequence for synthesize &  (\%) \\
\midrule
\midrule
Natural & - & 18.0  \\
Supervised & text\textsuperscript{$\dagger$} & 19.2  \\
\midrule
\multirow{2}{*}{Unsupervised} & text\textsuperscript{$\dagger$} & 21.7  \\
& ASR transcription\textsuperscript{$\ddagger$}  & 22.0  \\
\bottomrule
\end{tabular}
\begin{tablenotes}
\item{$\dagger$} \footnotesize{\text{Standard testing scenario.}}
\item{$\ddagger$} \footnotesize{\text{~For investigating the mismatch problem during inference only.}}
\item{~} \footnotesize{~~Obtained via decoding real audio with unsupervised ASR.}
\end{tablenotes}
\end{threeparttable}
\vspace{-12pt}
\end{table}

\noindent\textbf{Intelligibility:}
Results of evaluating intelligibility using the Google ASR API are listed in Table~\ref{tab:lj_wer}.
Speech from the unsupervised model recorded a 21.7\% WER, which is 2.5\% and 3.7\% higher than the supervised topline and the real audio.
This suggests that the unsupervised TTS achieved by the proposed method can be further improved.

Note that there is a mismatch between training and testing for the unsupervised model since it is trained on ASR transcription containing recognition errors and tested with phone sequence containing no error.
To investigate the impact of such a mismatch on the proposed method, we also measured the intelligibility of unsupervised TTS when the input is from an ASR transcription.
To be more specific, in the last row of Table~\ref{tab:lj_wer} we intentionally input phone sequences produced by unsupervised ASR (thus containing 6.97\% of recognition error as shown in Table~\ref{tab:lj_mos}).
Intuitively, inputting ASR transcriptions instead of real phone sequence should improve the performance since the mismatch problem can be considered solved.
Surprisingly, we found the intelligibility of the resulting speech did not improve.
This phenomenon suggests that the proposed unsupervised TTS method is fault-tolerable against first-stage ASR errors.


\vspace{6pt}
\noindent\textbf{Preference test against supervised TTS:} 
To better assess the value of unsupervised TTS, a pairwise comparison against the supervised model is performed, and the result is shown in Table~\ref{tab:lj_ab}.
In this test, listeners are asked to select a preference from the same sentence synthesized by the two different TTS models according to the naturalness or intelligibility.

\begin{table}[h]
\centering
\caption{A/B testing results comparing unsupervised method against the supervised counterpart on LJSpeech test split.
\label{tab:lj_ab}
}
\begin{tabular}[t]{ccc}
\toprule
& \multicolumn{2}{c}{Preference over Supervised }  \\
 \cmidrule(lr){2-3}
 & Naturalness & Intelligibility \\
\midrule
Unsupervised & 50.2\% & 54.0\% \\
\bottomrule
\end{tabular}
\vspace{5pt}
\end{table}

When comparing audio samples side-by-side, no obvious preference is observed in terms of naturalness.
Unexpectedly, the unsupervised model is preferred when testees are asked to make a judgment based on intelligibility.
While this suggests that the unsupervised model is indeed matching the performance of its supervised counterpart, we would also like to point out that the ground truth text is not available in the test, which might be the reason why the result differs from the ASR test.
Nevertheless, the A/B test shows that the unsupervised method is not worse than the supervised baseline.

\subsection{Multi-speaker TTS on LibriTTS}

\begin{table}[t!]
\vspace{-4pt}
\centering
\caption{Results on LibriTTS averaged over 6 different speakers. Naturalness measured by  MOS with 5-point scale. Intelligibility measured by WER with Google ASR.
\label{tab:lj_mos}}
\begin{tabular}[t]{lcc}
\toprule
{Method} & {MOS} & WER (\%) \\
\midrule
\midrule
Natural & 4.00 $\pm$ 0.07 & 23.4 \\
Supervised &  3.81 $\pm$ 0.08  & 25.3 \\
\midrule
Unsupervised &  3.71 $\pm$ 0.07  & 30.3 \\
\bottomrule
\end{tabular}
\end{table}

\noindent\textbf{Naturalness:}
With a similar score obtained by the real speech from the dataset, the score from the supervised TTS model reflects the challenge faced when switching to a multi-speaker setup.
Despite the gap between supervised and unsupervised TTS being larger than results on LJSpeech, human evaluations still suggest that the unsupervised TTS can still generate considerably realistic speech under this setup.

\vspace{6pt}

\noindent\textbf{Intelligibility:}
Unlike the single speaker setup, the WER gap between supervised and unsupervised TTS is larger, while the gap between supervised TTS and real speech remains similar.
Considering the phone error rate of pseudo-labels in the multi-speaker setup (7.78\%) is only slightly higher than for single speaker (6.97\%), the increasing WER reveals that modeling speech includes both speaker variance and annotation errors simultaneously, and remains a challenge for unsupervised TTS.

Given the above results, we conclude that the multi-speaker modeling problem in supervised TTS can potentially be even more difficult for the unsupervised setup.

\section{Conclusions}

We described the first framework for completely unsupervised speech synthesis.
The framework relies on the recent work in unsupervised speech recognition and the relatively mature neural-based speech synthesis paradigm.
As a preliminary study, we showed that the proposed TTS system can match the performance of supervised systems on an English dataset without using human annotation.
Future directions include better multi-speaker modeling, further reducing the resource requirement (for example, the need for lexicon), and experimenting on low resource languages for which unsupervised methods are more applicable.

\vspace{6pt}
\noindent\textbf{Acknowledgements} 
We thank Tomoki Hayashi and Erica Cooper for their advice on TTS training and evaluation.  This research was supported in part by the MIT-IBM Watson AI Lab.

\newpage
\bibliographystyle{IEEEtran}

\bibliography{mybib}

\begin{thebibliography}{10}
\providecommand{\url}[1]{#1}
\csname url@samestyle\endcsname
\providecommand{\newblock}{\relax}
\providecommand{\bibinfo}[2]{#2}
\providecommand{\BIBentrySTDinterwordspacing}{\spaceskip=0pt\relax}
\providecommand{\BIBentryALTinterwordstretchfactor}{4}
\providecommand{\BIBentryALTinterwordspacing}{\spaceskip=\fontdimen2\font plus
\BIBentryALTinterwordstretchfactor\fontdimen3\font minus
  \fontdimen4\font\relax}
\providecommand{\BIBforeignlanguage}[2]{{%
\expandafter\ifx\csname l@#1\endcsname\relax
\typeout{** WARNING: IEEEtran.bst: No hyphenation pattern has been}%
\typeout{** loaded for the language `#1'. Using the pattern for}%
\typeout{** the default language instead.}%
\else
\language=\csname l@#1\endcsname
\fi
#2}}
\providecommand{\BIBdecl}{\relax}
\BIBdecl

\bibitem{shen2018natural}
J.~Shen, R.~Pang, R.~J. Weiss, M.~Schuster, N.~Jaitly, Z.~Yang, Z.~Chen,
  Y.~Zhang, Y.~Wang, R.~Skerrv-Ryan \emph{et~al.}, ``Natural tts synthesis by
  conditioning wavenet on mel spectrogram predictions,'' in \emph{2018 IEEE
  international conference on acoustics, speech and signal processing
  (ICASSP)}.\hskip 1em plus 0.5em minus 0.4em\relax IEEE, 2018, pp. 4779--4783.

\bibitem{chung2019semi}
Y.-A. Chung, Y.~Wang, W.-N. Hsu, Y.~Zhang, and R.~Skerry-Ryan,
  ``Semi-supervised training for improving data efficiency in end-to-end speech
  synthesis,'' in \emph{ICASSP 2019-2019 IEEE International Conference on
  Acoustics, Speech and Signal Processing (ICASSP)}.\hskip 1em plus 0.5em minus
  0.4em\relax IEEE, 2019, pp. 6940--6944.

\bibitem{tan2021survey}
X.~Tan, T.~Qin, F.~Soong, and T.-Y. Liu, ``A survey on neural speech
  synthesis,'' \emph{arXiv preprint arXiv:2106.15561}, 2021.

\bibitem{baevski2021unsupervised}
\BIBentryALTinterwordspacing
A.~Baevski, W.-N. Hsu, A.~Conneau, and M.~Auli, ``Unsupervised speech
  recognition,'' in \emph{Advances in Neural Information Processing Systems},
  A.~Beygelzimer, Y.~Dauphin, P.~Liang, and J.~W. Vaughan, Eds., 2021.
  [Online]. Available: \url{https://openreview.net/forum?id=QmxFsofRvW9}
\BIBentrySTDinterwordspacing

\bibitem{liu2022towards}
A.~H. Liu, W.-N. Hsu, M.~Auli, and A.~Baevski, ``Towards end-to-end
  unsupervised speech recognition,'' \emph{arXiv preprint arXiv:2204.02492},
  2022.

\bibitem{wang2017tacotron}
Y.~Wang, R.~Skerry-Ryan, D.~Stanton, Y.~Wu, R.~J. Weiss, N.~Jaitly, Z.~Yang,
  Y.~Xiao, Z.~Chen, S.~Bengio \emph{et~al.}, ``Tacotron: Towards end-to-end
  speech synthesis,'' \emph{arXiv preprint arXiv:1703.10135}, 2017.

\bibitem{tachibana2018efficiently}
H.~Tachibana, K.~Uenoyama, and S.~Aihara, ``Efficiently trainable
  text-to-speech system based on deep convolutional networks with guided
  attention,'' in \emph{2018 IEEE International Conference on Acoustics, Speech
  and Signal Processing (ICASSP)}.\hskip 1em plus 0.5em minus 0.4em\relax IEEE,
  2018, pp. 4784--4788.

\bibitem{li2019neural}
N.~Li, S.~Liu, Y.~Liu, S.~Zhao, and M.~Liu, ``Neural speech synthesis with
  transformer network,'' in \emph{Proceedings of the AAAI Conference on
  Artificial Intelligence}, vol.~33, no.~01, 2019, pp. 6706--6713.

\bibitem{yamamoto2020parallel}
R.~Yamamoto, E.~Song, and J.-M. Kim, ``Parallel wavegan: A fast waveform
  generation model based on generative adversarial networks with
  multi-resolution spectrogram,'' in \emph{ICASSP 2020-2020 IEEE International
  Conference on Acoustics, Speech and Signal Processing (ICASSP)}.\hskip 1em
  plus 0.5em minus 0.4em\relax IEEE, 2020, pp. 6199--6203.

\bibitem{van2016wavenet}
A.~Van Den~Oord, S.~Dieleman, H.~Zen, K.~Simonyan, O.~Vinyals, A.~Graves,
  N.~Kalchbrenner, A.~W. Senior, and K.~Kavukcuoglu, ``Wavenet: A generative
  model for raw audio.'' \emph{SSW}, vol. 125, p.~2, 2016.

\bibitem{wang2015word}
P.~Wang, Y.~Qian, F.~K. Soong, L.~He, and H.~Zhao, ``Word embedding for
  recurrent neural network based tts synthesis,'' in \emph{2015 IEEE
  International Conference on Acoustics, Speech and Signal Processing
  (ICASSP)}.\hskip 1em plus 0.5em minus 0.4em\relax IEEE, 2015, pp. 4879--4883.

\bibitem{jia2021png}
Y.~Jia, H.~Zen, J.~Shen, Y.~Zhang, and Y.~Wu, ``Png bert: augmented bert on
  phonemes and graphemes for neural tts,'' \emph{arXiv preprint
  arXiv:2103.15060}, 2021.

\bibitem{tjandra2017listening}
A.~Tjandra, S.~Sakti, and S.~Nakamura, ``Listening while speaking: Speech chain
  by deep learning,'' in \emph{2017 IEEE Automatic Speech Recognition and
  Understanding Workshop (ASRU)}.\hskip 1em plus 0.5em minus 0.4em\relax IEEE,
  2017, pp. 301--308.

\bibitem{ren2019almost}
Y.~Ren, X.~Tan, T.~Qin, S.~Zhao, Z.~Zhao, and T.-Y. Liu, ``Almost unsupervised
  text to speech and automatic speech recognition,'' in \emph{International
  Conference on Machine Learning}.\hskip 1em plus 0.5em minus 0.4em\relax PMLR,
  2019, pp. 5410--5419.

\bibitem{liu2020towards}
A.~H. Liu, T.~Tu, H.-y. Lee, and L.-s. Lee, ``Towards unsupervised speech
  recognition and synthesis with quantized speech representation learning,'' in
  \emph{ICASSP 2020-2020 IEEE International Conference on Acoustics, Speech and
  Signal Processing (ICASSP)}.\hskip 1em plus 0.5em minus 0.4em\relax IEEE,
  2020, pp. 7259--7263.

\bibitem{baevski2020wav}
A.~Baevski, Y.~Zhou, A.~Mohamed, and M.~Auli, ``wav2vec 2.0: {A} framework for
  self-supervised learning of speech representations,'' in \emph{Proc. of
  NeurIPS}, 2020.

\bibitem{ljspeech17}
K.~Ito, ``The lj speech dataset,''
  \url{https://keithito.com/LJ-Speech-Dataset/}, 2017.

\bibitem{zen2019libritts}
H.~Zen, V.~Dang, R.~Clark, Y.~Zhang, R.~J. Weiss, Y.~Jia, Z.~Chen, and Y.~Wu,
  ``Libritts: A corpus derived from librispeech for text-to-speech,''
  \emph{arXiv preprint arXiv:1904.02882}, 2019.

\bibitem{panayotov2015librispeech}
V.~Panayotov, G.~Chen, D.~Povey, and S.~Khudanpur, ``Librispeech: an asr corpus
  based on public domain audio books,'' in \emph{Proc. of ICASSP}.\hskip 1em
  plus 0.5em minus 0.4em\relax IEEE, 2015, pp. 5206--5210.

\bibitem{g2pE2019}
K.~Park and J.~Kim, ``g2pe,'' \url{https://github.com/Kyubyong/g2p}, 2019.

\bibitem{gulrajani2017improved}
I.~Gulrajani, F.~Ahmed, M.~Arjovsky, V.~Dumoulin, and A.~C. Courville,
  ``Improved training of wasserstein gans,'' \emph{Advances in neural
  information processing systems}, vol.~30, 2017.

\bibitem{ott2019fairseq}
M.~Ott, S.~Edunov, A.~Baevski, A.~Fan, S.~Gross, N.~Ng, D.~Grangier, and
  M.~Auli, ``fairseq: A fast, extensible toolkit for sequence modeling,'' in
  \emph{Proceedings of NAACL-HLT 2019: Demonstrations}, 2019.

\bibitem{mohri2002weighted}
M.~Mohri, F.~Pereira, and M.~Riley, ``Weighted finite-state transducers in
  speech recognition,'' \emph{Computer Speech \& Language}, vol.~16, no.~1, pp.
  69--88, 2002.

\bibitem{hayashi2020espnet}
T.~Hayashi, R.~Yamamoto, K.~Inoue, T.~Yoshimura, S.~Watanabe, T.~Toda,
  K.~Takeda, Y.~Zhang, and X.~Tan, ``{Espnet-TTS}: Unified, reproducible, and
  integratable open source end-to-end text-to-speech toolkit,'' in
  \emph{Proceedings of IEEE International Conference on Acoustics, Speech and
  Signal Processing (ICASSP)}.\hskip 1em plus 0.5em minus 0.4em\relax IEEE,
  2020, pp. 7654--7658.

\bibitem{ren2020fastspeech}
Y.~Ren, C.~Hu, X.~Tan, T.~Qin, S.~Zhao, Z.~Zhao, and T.-Y. Liu, ``Fastspeech 2:
  Fast and high-quality end-to-end text to speech,'' \emph{arXiv preprint
  arXiv:2006.04558}, 2020.

\bibitem{jia2018transfer}
Y.~Jia, Y.~Zhang, R.~Weiss, Q.~Wang, J.~Shen, F.~Ren, P.~Nguyen, R.~Pang,
  I.~Lopez~Moreno, Y.~Wu \emph{et~al.}, ``Transfer learning from speaker
  verification to multispeaker text-to-speech synthesis,'' \emph{Advances in
  neural information processing systems}, vol.~31, 2018.

\bibitem{cooper2020zero}
E.~Cooper, C.-I. Lai, Y.~Yasuda, F.~Fang, X.~Wang, N.~Chen, and J.~Yamagishi,
  ``Zero-shot multi-speaker text-to-speech with state-of-the-art neural speaker
  embeddings,'' in \emph{ICASSP 2020-2020 IEEE International Conference on
  Acoustics, Speech and Signal Processing (ICASSP)}.\hskip 1em plus 0.5em minus
  0.4em\relax IEEE, 2020, pp. 6184--6188.

\bibitem{xue2022ecapa}
J.~Xue, Y.~Deng, Y.~Li, J.~Sun, and J.~Liang, ``Ecapa-tdnn for multi-speaker
  text-to-speech synthesis,'' \emph{arXiv preprint arXiv:2203.10473}, 2022.

\bibitem{desplanques2020ecapa}
B.~Desplanques, J.~Thienpondt, and K.~Demuynck, ``Ecapa-tdnn: Emphasized
  channel attention, propagation and aggregation in tdnn based speaker
  verification,'' \emph{arXiv preprint arXiv:2005.07143}, 2020.

\bibitem{ravanelli2021speechbrain}
M.~Ravanelli, T.~Parcollet, P.~Plantinga, A.~Rouhe, S.~Cornell, L.~Lugosch,
  C.~Subakan, N.~Dawalatabad, A.~Heba, J.~Zhong \emph{et~al.}, ``Speechbrain: A
  general-purpose speech toolkit,'' \emph{arXiv preprint arXiv:2106.04624},
  2021.

\bibitem{wang2018style}
Y.~Wang, D.~Stanton, Y.~Zhang, R.-S. Ryan, E.~Battenberg, J.~Shor, Y.~Xiao,
  Y.~Jia, F.~Ren, and R.~A. Saurous, ``Style tokens: Unsupervised style
  modeling, control and transfer in end-to-end speech synthesis,'' in
  \emph{International Conference on Machine Learning}.\hskip 1em plus 0.5em
  minus 0.4em\relax PMLR, 2018, pp. 5180--5189.

\bibitem{kong2020hifi}
J.~Kong, J.~Kim, and J.~Bae, ``Hifi-gan: Generative adversarial networks for
  efficient and high fidelity speech synthesis,'' \emph{Advances in Neural
  Information Processing Systems}, vol.~33, pp. 17\,022--17\,033, 2020.

\bibitem{kumar2019melgan}
K.~Kumar, R.~Kumar, T.~de~Boissiere, L.~Gestin, W.~Z. Teoh, J.~Sotelo,
  A.~de~Br{\'e}bisson, Y.~Bengio, and A.~C. Courville, ``Melgan: Generative
  adversarial networks for conditional waveform synthesis,'' \emph{Advances in
  neural information processing systems}, vol.~32, 2019.

\end{thebibliography}

\end{document}